\documentclass[aps,showpacs, nofootinbib]{revtex4}

\newcommand{\be}{\begin{equation}}
\newcommand{\ee}{\end{equation}}
\newcommand{\ben}{\begin{eqnarray}}
\newcommand{\een}{\end{eqnarray}}

\newcommand{\la}{{\lambda}}

\newcommand{\cO}{{\cal O}}

\newcommand{\cL}{{\cal L}}

\newcommand{\p}{\partial}
\newcommand{\na}{\nabla}

\newcommand{\tcF}{{\tilde {\cal F}}}
\newcommand{\tcA}{{\tilde {\cal A}}}

\newcommand{\tF}{\tilde F}
\newcommand{\tE}{\tilde E}
\newcommand{\tB}{\tilde B}

\newcommand{\tla}{\tilde {\lambda}}

\newcommand{\ep}{\epsilon}

\newcommand{\bla}{\bar \lambda}

\newcommand{\ga}{\gamma}


\begin{document}

\title{Complex scalar field in strictly stationary Einstein-Maxwell-axion-dilaton 
spacetime with negative cosmological constant}

\author{Bartlomiej Bakon and Marek Rogatko}
\affiliation{Institute of Physics \protect \\
Maria Curie-Sklodowska University \protect \\
20-031 Lublin, pl.~Marii Curie-Sklodowskiej 1, Poland \protect \\
marek.rogatko@poczta.umcs.lublin.pl \protect \\
rogat@kft.umcs.lublin.pl}

\date{\today}

\begin{abstract}
We proved that strictly stationary Einstein-Maxwell-axion-dilaton spacetime
with negative cosmological constant could not support a nontrivial configurations of
complex scalar fields. We considered the general case of the arbitrary number of $U(1)$-gauge fields
in the theory under consideration.
 
\end{abstract}

\maketitle

\section{Introduction}
One of the most important issues of general relativity or its generalization to higher dimensions are 
connected with a gravitational collapse
and the emergence of black holes or black objects.
Recently, self-gravitating objects and gravitational collapse
in asymptotically anti-de Sitter (AdS) spacetime gain much attention. 
One of the reason is that there is a surprising connection between general relativity and
non-gravitational fields of physics. An exact correspondence between gravity theory in $(d+1)$-dimensional
AdS spacetime and a conformal field theory of the boundary of this spacetime was proposed \cite{adscft}.
The AdS/CFT correspondence becomes a very useful tool which helps us to understand
the strong coupled gauge theories as well as it has been applied to the condensed matter physics,
including superconductivity and superfluidity.\\
In Ref.\cite{bou84} it was shown that the only strictly stationary asymptotically
AdS spacetime fulfilling  the vacuum Einstein equations with negative cosmological constant
is the AdS spacetime. Uniqueness of AdS spacetime in higher dimensions was studied in \cite{qin04}, where
it was also revealed that the vacuum solution of Einstein equations with a negative cosmological constant
is AdS spacetime. The global aspects of the complete nonsingular asymptotically
locally AdS spacetime being the solution of Einstein vacuum equations were elaborated in Ref.\cite{and06}.
On the other hand, the properties and uniqueness of AdS solitons were considered in \cite{gal},
while the problem of non-existence of black holes under certain asymptotic conditions was tackled in \cite{gal03}.
The systematic classification of static vacuum spacetimes with negative cosmological constant was conducted
in Ref.\cite{chr01}.
\par
A different issue, related to the problem of
gravitational collapse in generalization of Einstein theory to higher dimensions and
emergence of higher dimensional black objects (like black rings, black Saturns) and 
multidimensional black holes
was widely studied. The complete classification of $n$-dimensional 
charged black holes both with non-degenerate and 
degenerate component of the event horizon was proposed in Refs.\cite{nd}, while
partial results for the very nontrivial case of $n$-dimensional rotating black hole uniqueness theorem were 
provided in \cite{nrot}. These researches comprise also the case of the low-energy limit 
of the string theory, like dilaton gravity, Einstein-Maxwell-axion-dilaton (EMAD)-gravity and 
supergravities theories \cite{sugra}.
On the other hand, black holes and 
their properties as key ingredients of the AdS/CFT attitude to superconductivity
also acquire great attention. 
The nonlinear evolution of a weakly perturbed AdS spacetime by solving numerically spherically symmetric
Einstein massless scalar field was studied in \cite{biz11}. It was shown that AdS spacetime
is unstable under arbitrary small generic perturbations, both in four and higher dimensions \cite{jal11}.
Recently, the numerical solutions of black holes with only one Killing
vector were presented \cite{dia11}. This investigation was motivated by the superradiance arguments
\cite{kun06} that AdS black hole might exist having only one Killing vector. Moreover,
topological defects like vortex \cite{deh02} or interaction 
domain walls with AdS black holes \cite{mod06} were also taken into account. These works naturally pose a 
question about possible configurations in AdS spacetime with certain matter fields. In Ref.\cite{shi12}
it was revealed that strictly stationary AdS spacetime could not allow for the existence of nontrivial
configurations of complex scalar fields or form fields.
\par
Motivated by the aforementioned problems of strictly stationary AdS spacetimes with matter as well as
the wide range of possible applications of the AdS/CFT correspondence, we shall investigate
the problem of possible matter configurations in EMAD-gravity.
EMAD-gravity is the low-energy limit of heterotic string compactified to four-dimensions. For the generality
of the researches we shall not restrict our consideration to only one gauge field. One takes into account the 
arbitrary number of $U(1)$-gauge fields. In some sense we generalize considerations conducted on 
Eistein-Maxwell (EM) background \cite{shi12} to the more complicated gravity theory.\\
Our paper is organized as follows. In Sec.II we describe 
theory under consideration in terms of $U(1)$-gauge field strengths and their $SL(2,~R)$-duals.
Then, using the rigid positive energy theorem \cite{wit}, we find that strictly stationary
EMAD spacetime with negative cosmological constant cannot be nontrivial configurations of
complex scalar field. Then, one concludes the investigations.

\section{EMAD-gravity}
Motivated by the recent works concerning the numerical solutions of Einstein complex scalar field systems
we shall consider the generalization of Einstein-Maxwell (EM) complex scalar field system \cite{dia11} 
to the low-energy limit of the heterotic string theory, the so-called Einstein-Maxwell-axion-dilaton gravity (EMAD) with
cosmological constant. Moreover, we take into account complex scalar field existing on this background.
The theory under consideration will contain gravitation field $g_{\mu \nu}$, arbitrary number $N$ of
$U(1)$-gauge fields, the dilaton field $\phi$ and axion $a$. The action will be provided by \cite{kal}
\ben \label{emad}
S = \int d^4 x~\sqrt{-g}~\bigg[
R &-& 2~\Lambda - 2~\na_\mu \phi \na^\mu \phi - {1 \over 2}~\na_\mu a~\na^\mu a -
\sum \limits_{n=1}^N e^{-2 \phi}~F_{\mu \nu}^{(n)}~F^{\mu \nu ~(n)}
- \sum \limits_{n=1}^N a~F_{\mu \nu}^{(n)}~\ast F^{\mu \nu ~(n)} \\ \nonumber
&-& 2~\mid \na~\pi \mid^2 
\bigg],
\een
where we have denoted the strength of the adequate 
gauge field $F_{\mu \nu}^{(n)} = 2~\na_{[ \mu}A_{\nu ]}^{(n)}$. Its dual
is given by 
$\ast~F_{\mu \nu}^{(n)} = {1 \over 2}~\ep_{\mu \nu \alpha \beta}F^{\mu \nu~ (n)}$ and by
$\pi$ we denote a complex scalar field. 
One should remark that when the number of vector fields is six we obtain $N=4,~d=4$ bosonic part of supergravity
theory.
For the sake of generality one will keep the number of $U(1)$-gauge fields $n$ to be arbitrary.
\par
In many physical problems \cite{kal} the action describing by the relation (\ref{emad}) can be written
in a more convenient form.
Namely, introducing
a complex scalar {\it axi-dilaton} which implies
\be
\la = a + i~e^{-2 \phi},
\ee
and defining $SL(2,~R)$-duals to the gauge fields $F_{\mu \nu}^{(n)}$, the action in question 
yields
\be
S = \int d^4 x~\sqrt{-g}~\bigg[
R - 2~\Lambda + 2~{\na_\mu \la~\na^\mu \tla \over (\la -\tla)^2}
+ \sum\limits_{n=1}^N F_{\mu \nu}^{(n)}~\ast \tcF^{\mu \nu~(n)}
- ~\mid \na~\pi \mid^2
\bigg],
\ee
where the $SL(2,~R)$-duals are provided by the relation
\be
\tcF_{\alpha \beta}^{(n)} = e^{-2 \phi}~\ast F_{\alpha \beta}^{(n)} - a~F_{\alpha \beta}^{(n)}.
\ee
The equation of motion for $SL(2,~R)$-duals is of the form
$\na_\alpha~\ast \tcF^{\alpha \beta~(n)} = 0$ which entails the existence of $N$ vector potentials
$\tcA_{\beta}^{(n)}$ satisfying relation of the form
\be
\tcF_{\alpha \beta}^{(n)} = 2~\na_{[ \alpha} \tcA^{(n)}_{\beta ]}.
\ee 
Consequently, the analogous relation for $F_{\mu \nu}^{(n)} = 2~\na_{[ \mu} A^{(n)}_{\nu ]}$
is not a consequence of equations of motion but it stems from the Bianchi identity.\\
The energy-momentum tensor components in the theory in question $T_{\alpha \beta}(F,~\tcF,~\la,~\Pi)
= T_{\alpha \beta}(\Pi) + T_{\alpha \beta}(F,~\tcF,~\la)$ are given respectively
\be
T_{\alpha \beta}(\pi) = -2~g_{\alpha \beta}~\na_\ga \pi~\na^\ga \pi^\ast + 2~\bigg(
\na_{\alpha} \pi~\na_\beta \pi^\ast + \na_{\alpha} \pi^\ast~\na_\beta \pi \bigg),
\ee
for the complex scalar field, while for the $U(1)$-gauge fields may be written in the form as follows:
\be
T_{\alpha \beta}(F,~\tF,~\la) =
- \bigg[ 4~\sum\limits_{n=1}^N F_{\alpha \delta~(n)}~\ast \tcF_{\beta}{}^{\delta~(n)}
- g_{\alpha \beta}~\sum\limits_{n=1}^N F_{\alpha \beta}^{(n)}~\ast \tcF^{\alpha \beta ~(n)} \bigg]
+
{2~g_{\alpha \beta}~\na_{\ga}\la \na^\ga \bla - 4~\na_\alpha \la \na_\beta \bla \over
(\la - \bla)^2}.
\ee
In our paper we shall examine the strictly stationary spacetime, which is not restricted to the axisymmetric
ones and does not contain any black hole.
The strictly stationary spacetime by the definition posses everywhere
a timelike Killing vector field $k_{\delta}$.
Due to the strictly stationarity, one assumes that all the field in the considered theory will be stationary.
Namely, one has that
\be
\cL_k~F_{\alpha \beta} = \cL_k~\tcF_{\alpha \beta} = \cL_k~a = \cL_k~\phi = \cL_k~\Pi = 0.
\ee
It can be proved, having in mind the exact form of the energy-momentum tensor $T_{\alpha \beta}(F,~\tcF,~\la,~\Pi)$
that it will satisfy the stationarity assumption, i.e., 
$\cL_k~T_{\alpha \beta}(F,~\tcF,~\la,~\Pi) = 0$.\\
Furthermore one can define the twist vector $\omega_a$ of stationary Killing vector field $k_a$ by the
following definition:
\be
\omega_a = {1 \over 2}~\ep_{abcd}~k^b~\na^c~k^d.
\ee
Using the formula valid for any Killing vector field 
$\na_\alpha~\na_\beta \chi_\ga = - R_{\beta \ga \alpha}{}{}^{\delta}~\chi_\delta$, and carrying
out the differentiation of the twist vector, one leads to the identity satisfied by the stationary
Killing vector field $k_\alpha$
\be
\na_\beta ~\omega_\alpha = {1 \over 2}~\ep_{\alpha \beta \ga \delta}~k^{\ga}~R^{\delta \chi}~k_{\chi},
\label{rr}
\ee
By virtue of the definition of $\omega_\alpha$ it follows directly that the following yields:
\be
\na_\alpha~\bigg( {\omega^\alpha \over V^4} \bigg) = 0,
\ee
where one denotes $V^2 = - k_\ga~k^\ga$.\\
To proceed further,
we define electric and magnetic components for gauge field strengths $F_{\alpha \beta}^{(n)}$
and $\tcF_{\alpha \beta}^{(n)}$. Namely, for electric components we have
\be
E_{\alpha}^{(n)} = - F_{\alpha \beta}^{(n)}~k^\beta, \qquad
\tE_{\alpha}^{(n)} = - \tcF_{\alpha \beta}^{(n)}~k^\beta,
\ee
while magnetic ones are provided by
\be
B_{\alpha}^{(n)} = {1 \over 2}~\ep_{\alpha \beta \ga \delta}~k^\beta~F^{\ga \delta~(n)}, \qquad
\tB_{\alpha}^{(n)} = {1 \over 2}~\ep_{\alpha \beta \ga \delta}~k^\beta~\tcF^{\ga \delta~(n)}.
\ee
Consequently, in terms of the above magnetic and electric components, 
the field strength $F_{\alpha \delta}^{(n)}$ is given by the following relation:
\be
V^2~F_{\alpha \beta}^{(n)} = - 2~k_{[\alpha} E_{\beta]}^{(n)} + 
\ep_{\alpha \beta \ga \delta}~k^\ga~B^{\delta~(n)},
\ee
while for $SL(2,~R)$-duals, one obtains that $\tcF_{\alpha \delta}^{(n)}$ yields
\be
V^2~\tcF_{\alpha \beta}^{(n)} = - 2~k_{[\alpha} \tE_{\beta]}^{(n)} + 
\ep_{\alpha \beta \ga \delta}~k^\ga~\tB^{\delta~(n)}.
\ee
On the other hand, equations of motion
for magnetic and electric parts for each gauge field strength imply
\ben
\na_{\alpha}\bigg( {\tE^{\alpha~(n)} \over V^2 } \bigg) &=& 2~{\tB^{\ga~(n)} \over V^2}~\omega_\ga,\\
\na_{\alpha}\bigg( {E^{\alpha~(n)} \over V^2} \bigg) &=& 2~{B^{\ga~(n)} \over V^2}~\omega_\ga,\\
\na_{\alpha}\bigg( {\tB^{\alpha~(n)} \over V^2} \bigg) &=& - 2~{\tE^{\ga~(n)} \over V^2}~\omega_\ga,\\
\na_{\alpha}\bigg( {B^{\alpha~(n)} \over V^2} \bigg) &=& - 2~{E^{\ga~(n)} \over V^2}~\omega_\ga.
\een
By virtue of the field invariance conditions
$\cL_k~F_{\alpha \beta}^{(n)} = \cL_k~\tcF_{\alpha \beta}^{(n)} = 0$, as well as
relations $\na_{[ \ga} F_{\alpha \beta ]}^{(n)} = \na_{[ \ga }\tcF_{\alpha \beta ]}^{(n)} = 0$, 
one achieves the generalized Maxwell source-free equations. Their explicit forms are provided by
\ben
\na_{[\alpha}E_{\beta ]}^{(n)} &=& \na_{[\alpha}\tE_{\beta ]}^{(n)} = 0,\\
\na_{[\alpha}B_{\beta ]}^{(n)} &=& \na_{[\alpha}\tB_{\beta ]}^{(n)} = 0.
\een
The demand that the spacetime under consideration is simply connected, enables one to introduce the
adequate potential for each electric and magnetic components. They are obliged to satisfy
\ben
E_{\alpha}^{(n)} = \na_{\alpha} \varphi^{(n)}, \qquad
\tE_{\alpha}^{(n)} = \na_{\alpha} \Phi^{(n)},\\
B_{\alpha}^{(n)} = \na_{\alpha} \psi^{(n)}, \qquad
\tB_{\alpha}^{(n)} = \na_{\alpha} \Psi^{(n)}.
\een
Taking into account Eq.(\ref{rr}) and the explicit form
of the Ricci tensor for the fields in question, we can assert that the following equality takes place:
\be
\na_{[\alpha} \omega_{\beta ]} = 4~\sum \limits_{n=1}^N B_{[ \alpha}^{(n)}~\tB_{\beta ]}^{(n)}.
\label{oom}
\ee
The right-hand side of (\ref{oom}) depicts the {\it generalized Poynting flux} in EMAD-gravity.
\par
The fact that electric and magnetic part of the adequate gauge field strength can be expressed in
terms of the potentials, enables one to find other relations binding them with the aforementioned twist vector.
Thus, by the direct calculations it can be revealed that
one arrives at the relations binding together the twist vector, the potentials $\psi^{(n)}$
and $\tB_{\beta }^{(n)}$
\be
\na_{[ \alpha}~\bigg( \omega_{\beta ]} + 2~\sum \limits_{n=1}^N \psi^{(n)}~\tB_{\beta ]}^{(n)} \bigg) = 0,
\ee
and the other for $ \Psi^{(n)}$ and $B_{\beta }^{(n)} $
\be
\na_{[ \alpha}~\bigg( \omega_{\beta ]} - 2~\sum \limits_{n=1}^N \Psi^{(n)}~B_{\beta ]}^{(n)} \bigg) = 0.
\ee
Consequently, the existence of the following scalar functions are guaranteed
\be
\na_\alpha U_{B^{(n)}} = \omega_\alpha + 2~\sum \limits_{n=1}^N \psi^{(n)}~\tB_\alpha^{(n)},
\ee
\be
\na_\alpha U_{\tB^{(n)}} = \omega_\alpha - 2~\sum \limits_{n=1}^N \Psi^{(n)}~B_\alpha^{(n)}.
\label{pot}
\ee
Having in mind the generalized forms of Maxwell equations in the theory in question,
one is able to find the following relations fulfilling by $U_{B^{(n)}}$
\be
\na_{\alpha} ~\bigg(
U_{B^{(n)}}~{\omega^\alpha \over V^4} - \sum \limits_{n=1}^N {\psi^{(n)}~\tE^{\alpha (n)} \over V^2} 
\bigg) =
{\omega_\beta~\omega^\beta \over V^4} - \sum \limits_{n=1}^N {B_\alpha^{(n)}~\tE^{\alpha (n)} \over V^2},
\label{u1}
\ee
and by the scalar function $U_{\tB^{(n)}}$
\be
\na_{\alpha} ~\bigg(
U_{\tB^{(n)}} {\omega^\alpha \over V^4} + \sum \limits_{n=1}^N {\Psi^{(n)}~E^{\alpha (n)} \over V^2} 
\bigg)
=
{\omega_\beta~\omega^\beta \over V^4} + \sum \limits_{n=1}^N {\tB_\alpha^{(n)}~E^{\alpha (n)} \over V^2}.
\label{u2}
\ee
Further, combining the above expressions in order to find
$2~R_{\alpha \beta}~k^\alpha k^\beta /V^2$, one obtains the relation of the form as
\be
\na_{\alpha} \bigg(
{\na^\alpha V^2 \over V^2} + 4~{\omega_\alpha~\omega^\alpha \over V^4} \bigg)
= - 2~\Lambda + 2~\sum \limits_{n=1}^N{\tE_{\beta}^{(n)}~B^{\beta (n)} - ~E_{\beta}^{(n)}~\tB^{\beta (n)}
\over V^2}.
\label{uu}
\ee
Combining Eqs.(\ref{u1})-(\ref{u2}) and (\ref{uu}), 
by a direct calculation, 
we reach to the following
expression valid for each of the gauge fields in question:
\be
\na_\alpha \bigg( {\na^\alpha V^2 \over V^2} + \Theta^{\alpha~(n)} \bigg) = - 2~\Lambda,
\label{lam}
\ee
where we have denoted by $\Theta^{\alpha~(n)}$
\be
\Theta^{\alpha~(n)} = 2~\bigg( U_{B^{(n)}} + U_{\tB^{(n)}} \bigg)~{\omega^\alpha \over V^4}
+ 2~\sum \limits_{n=1}^N {(\Psi^{(n)}~E^{\alpha (n)} - \psi^{(n)}~\tE^{\alpha (n)}) \over V^4}.
\ee

\par
Considering the case when $\Lambda < 0$, one can
follow Ref.\cite{shi12} and introduce the vector field $r^\beta$, provided by
\be
\na_\ga~ r^\ga = - 2~\Lambda.
\label{ass}
\ee
As in Ref.\cite{bou84} one can define a spacetime which is AdS if there exists a chart $(t,~r,~\theta,~\phi)$
defined outside a spatial compact world tube. The line element of it is subject to the relation
\ben
ds^2 = ds_0^2 &+& \cO(r^{-2})~dt^2 + \cO(r^{-1})~dr^2 \\ \nonumber
&+& \cO(r)(remaining~ differentials~ not~ including~dr) \\ \nonumber
&+& \cO(r^{-1})(remaining~ differentials~ not~ including~dr),
\een
where $ds_0^2 $ is Schwarzschild-anti-de Sitter metric. On the other hand, near infinity $r^\alpha$
will imply
\be
r^\alpha \simeq \Lambda~r~\bigg( {\p \over \ r} \bigg)^\alpha + \dots
\ee
The global existence of  $r^\alpha$ is justified by the assumption that  $r^\alpha = \na^\alpha \chi$,
where on its turn $\chi$ fulfils the equation $\na^2 \chi = - 2~\Lambda$. Just, one can redefine
$\chi$ in order to subtracted the term $-2~\Lambda$ and obtain $\na^2 {\tilde \chi} = S$, where $S$
is a nonsingular source term. The existence of the solution of the above equation is guaranteed
on a regular Riemaniann manifold, so one can always introduce $r^\alpha$.
\par
It implies that Eq.(\ref{lam}) can be reorganize to the following form:
\be
\na_{\alpha} \bigg(
{\na^\alpha ~V^2 \over V^2} - r^\alpha + \Theta^{\alpha~(n)} \bigg) = 0.
\label{pot1}
\ee
The value of the left-hand side of the above integral can be rewritten in the form of the surface integral
proportional to the total mass $M$.\\
Having in mind the rigid positive energy theorem \cite{wit}, and taking into account the volume integral of relation
(\ref{pot1}) which it equal to zero, one concludes that $M = 0$. It yields that the spacetime under consideration
should be the exact anti-de-Sitter one. One remarks that we get no nontrivial
self-gravitating solutions for complex scalar fields in EMAD-gravity with negative cosmological
constant, under the assumptions imposed here, i.e., under the strictly stationarity of the considered spacetime.
\par
Summing up, we have revealed that the strictly stationary,
simply connected spacetime in EMAD-gravity with positive cosmological constant
have no nontrivial configurations of complex scalar fields. The spacetime
under consideration ought to reduce to be exactly Minkowski or anti-de Sitter depending on the
occurrence of negative cosmological constant.

\section{Conclusions}
In our paper we have considered the problem of the possible emergence of a self-gravitating structure
composed of complex scalar fields. We considered EMAD-gravity 
being the low-energy limit of heterotic string theory, which constitutes the generalization of EM theory with 
additional fields of axion and dilaton. One takes into account the case of negative cosmological constant.
For the sake of generality of our investigations, we elaborated the case of an arbitrary number of
$U(1)$-gauge fields. One restricts his consideration to the case when the spacetime in question 
is strictly stationary, i.e., if it admits nowhere vanishing
timelike Killing vector field. Rewriting the action describing the considered physical system 
in terms of $U(1)$-gauge strengths  and their $SL(2,~R)$-duals, enables us to implement the rigid positive 
energy theorem \cite{wit} and conclude that strictly stationary EMAD 
spacetime with negative cosmological constant could not support
nontrivial configurations of complex scalar fields. It means that, as well as
in the case of EM theory with negative cosmological constant, one comes to a conclusion
that a self-gravitating complex scalar fields cannot exist in the theory in question.
\par
Of course, it remains for the future investigations how to overcome the problem of getting rid
of the assumption (\ref{ass}). We hope to return to this problem elsewhere.


\begin{acknowledgments}
MR was partially supported by the grant of the National Science Center
$2011/01/B/ST2/00408$.
\end{acknowledgments}



\end{document}